\documentclass[%
reprint,
superscriptaddress,
footinbib,
bibnotes,
amsmath,amssymb,
balancelastpage,
flushbottom,
pra,
floatfix,
]{revtex4-2}

\usepackage{ulem}
\usepackage{dcolumn}
\usepackage{bm}

\usepackage[T1]{fontenc}
\usepackage[utf8]{inputenc}
\usepackage[english]{babel}
\usepackage[dvipsnames]{xcolor}
\usepackage[pdftex]{graphicx}
\usepackage[colorlinks,allcolors=blue]{hyperref}

\begin{document}
\author{M.\ V.\ Dubinin}
\author{A.\ A.\ Fomin} 
\author{G.\ G.\ Kozlov}
\author{M.~Yu.\ Petrov}
\author{V.~S.\ Zapasskii}
\affiliation{Spin Optics Laboratory, St.~Petersburg State University, 198504 St.~Petersburg, Russia}

\begin{abstract} 
We propose a simple method of measuring the autocorrelation function of a spin noise based on multiplication and averaging two digitized signal traces, with one of them being a time-reversed copy of the other. This procedure allows one to obtain, with lower computational expenses, all the information usually derived in the Fourier transform spin-noise spectroscopy, retaining all the merits of the latter.  We successfully applied this method to the measurements of spin noise in cesium vapors by using a digital oscilloscope in the capacity of the analog-to-digital converter. Specific opportunities of this experimental approach as applied to a more general problem of studying the nature of light-intensity noise are discussed. 
\end{abstract}

\title {Measuring spin-noise correlation function via time reversal}
\maketitle


The most powerful specialized methods of studying electron spin systems are based on the effect of magnetic resonance observed when the frequency of the applied ac magnetic field coincides with that of spin precession of the system in an external dc magnetic field. 
Among different methods of the electron paramagnetic resonance (EPR) spectroscopy, an important place is occupied by the spin-echo technique\ \cite{hahn} that allows one to perform effective time reversal in the system of precessing spins, thus making it possible to distinguish between the processes of reversible and irreversible dephasing and to realize a highly specific approach to the EPR spectroscopy. 
It is important that an indispensable feature of all methods of conventional EPR spectroscopy is the presence of coherent spin precession. 

In the last two decades, there has been developed a conceptually new method of magnetic-resonance detection that is based on the fluctuation-dissipation theorem\ \cite{weber,kubo} and does not imply resonant excitation of spin precession and restricts itself to detecting spontaneous noise of the spin-system magnetization\ \cite{alzap, crooker, zap-review}. 
This method referred to as the {\it spin noise spectroscopy} (SNS) turned at present into a fairly popular tool of research of spin-systems in atomic vapors, in semiconductors, and in dielectrics with paramagnetic impurities\ \cite{sn1, sn2, sn3, sn4, cronenberger15, GlasenappPRL14, sn5, sn6, sn7, sn8}. 
In contrast to all conventional methods of EPR spectroscopy, detection of spin resonance in the SNS does not imply any coherent precession and even any spin polarization. 
Under these conditions, amplitude measurements of regular signals become inapplicable, and the experimental protocols need to be modified.  
As a result, certain methods and effects of conventional EPR spectroscopy appear to be irreproducible in a straightforward way in the SNS. 
In particular, this is true with respect to the spin-echo effect that implies pulsed measurements in a coherently excited spin system. 
At the same time, the spin noise of a steady-state spin system (like any other time-dependent process) formally subjected to the time reversal should inevitably reproduce all the states of the process preceding the moment of reversal. 
In this paper, we show that this simple operation, in combination with joint processing of the direct and reverse data-arrays, may be highly useful for direct measurement of the noise correlation function. 

\begin{figure}[t]
\includegraphics[width=\columnwidth]{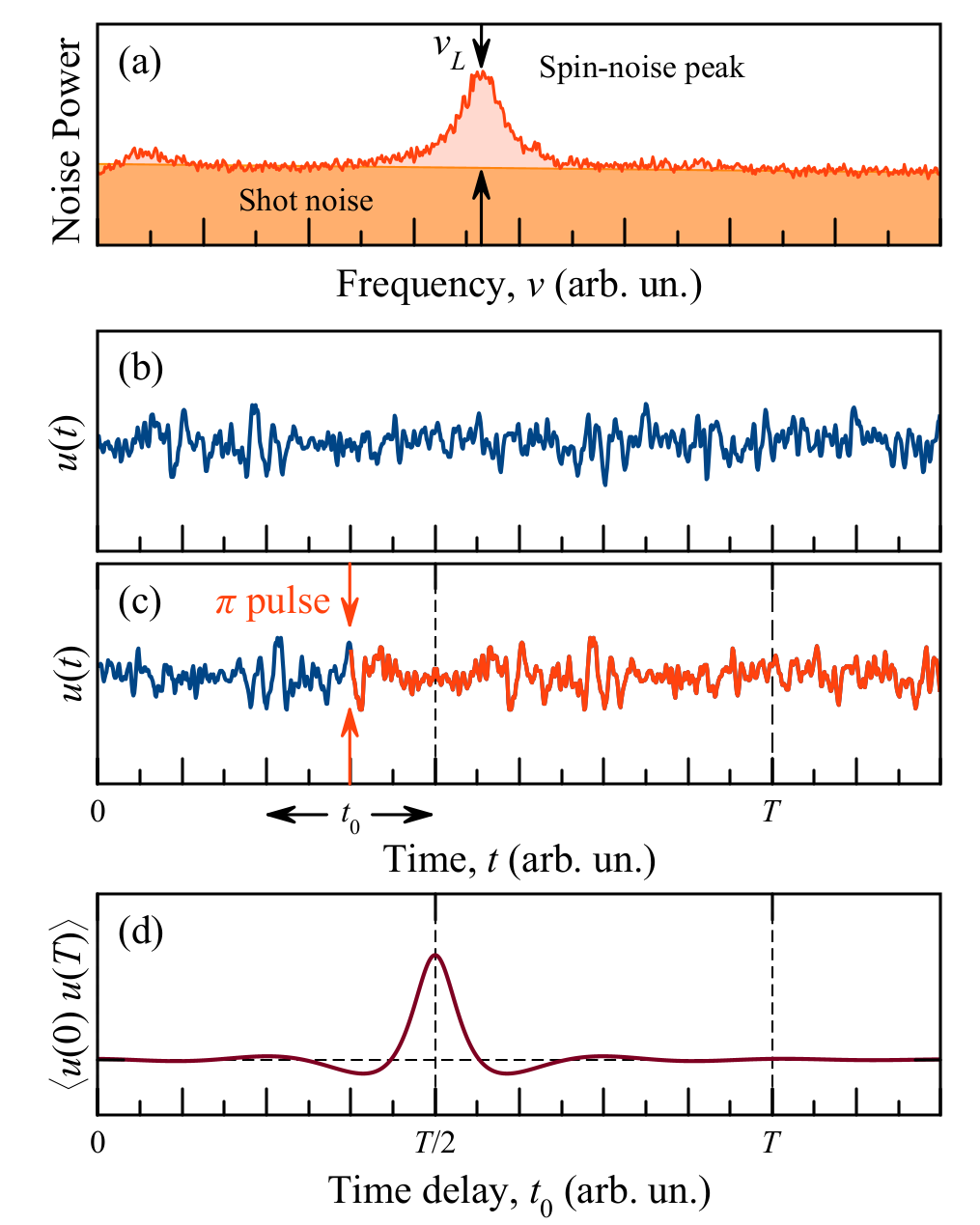}
\caption{Typical frequency (a) and time (b) dependences of signals detected in spin noise spectroscopy. (c) Illustration to the {\it gedanken} experiment on observation of spin-echo effect in spin noise.  The $\pi$-pulse is applied at $t = t_0$. (d) Expected dependence of cross-correlation of the noise signals at moments $0$ and $T$ as a function of $t_0$. 
}\label{fig1}
\end{figure}

In the conventional spin-noise measurements, one detects fluctuations of the polarization-plane rotation of a probe laser beam transmitted through (or reflected from) the sample. 
Generally, the experimentally detected noise comprises of two main contributions: the noise directly caused by spin motion (the spin-noise proper) and the shot noise. 
As a result, the spin-noise spectrum (in a transverse magnetic field) usually looks as a Lorentzian peak (centered at the Larmor frequency $\nu_L$) on the background of the flat (``white'') shot-noise spectrum [Fig.\ \ref{fig1}(a)]. 
Since the magnitude of the spin noise is usually smaller than that of the shot noise, the trace of the detected noise signal looks as shown in Fig.\ \ref{fig1}(b) (with no clearly seen harmonic signal of the spin precession). 

It should be also noted that, in conformity with the Wiener-Khinchin theorem\ \cite{wiener, khinchin}, the incoherent spin precession revealed in the spin-noise spectrum, can be no less efficiently detected in its autocorrelation function. 
At the same time, such an approach, which may be easier and cheapper in realization than the Fourier-transform-based method\ \cite{bartlett, welch}, practically is not used in the standard SNS.  
A few experimental works, that were aimed at measuring the spin-noise correlation function, were fulfilled with pulsed (mode-locked) lasers, with the inter-pulse spacing used for measuring and calibration of spin dynamics\ \cite{StarosielecAPL08,berski, pursley}.   

Here, we want  to attract attention to the fact that, in spite of absence of explicit coherent precession, the signal of the {\it incoherent} precession, in principle, can be used to observe the spin-echo effect.
Indeed, if the spin system is subjected, at time $t_0$, to a $\pi$-pulse of the resonant microwave field, then all elementary precessions change their sign thus giving rise to the effective time reversal in the spin system [Fig.\ \ref{fig1}(c)]. 
After this pulse, the spin noise signal evidently does not show any visible changes but starts to evolve in the {\it backward} direction, and the signals detected at moments $\pm\tau$ with respect to $t_0$ appear to be {\it correlated}. 
Thus, by measuring cross-correlation of the noise signals at moments $0$ and $T$ [Fig.\ \ref{fig1}(d)], as a function of $t_0$, we will observe the spin-echo signal. 
In this {\it gedanken} experiment, we will obtain, as can be expected, the spin-noise correlation function with its width controlled  by the irreversible dephasing of the precession rates $T_2$ and with its amplitude relaxing with the longitudinal relaxation time $T_1$. 
Note that a similar  effect of suppression of the reversible spin relaxation in spin noise was considered in Refs.\ \cite{braun,poshakin}, with the $\pi$ pulse replaced by the magnetic-field inversion.

In this work, we do not intend to perform this experiment and will not enter into technical details of its realization. 
The above  experiment is described just to demonstrate applicability of the effective time reversal for restoration of irregular signals using their correlation (rather than amplitude-related) characteristics. 
We propose here an imitation of the time reversal that retains certain useful properties of a real-time reversal and can be used, by means of a simple experimental procedure, to measure a correlation function of the spin noise. 

\begin{figure}[t]
\includegraphics[width=\columnwidth]{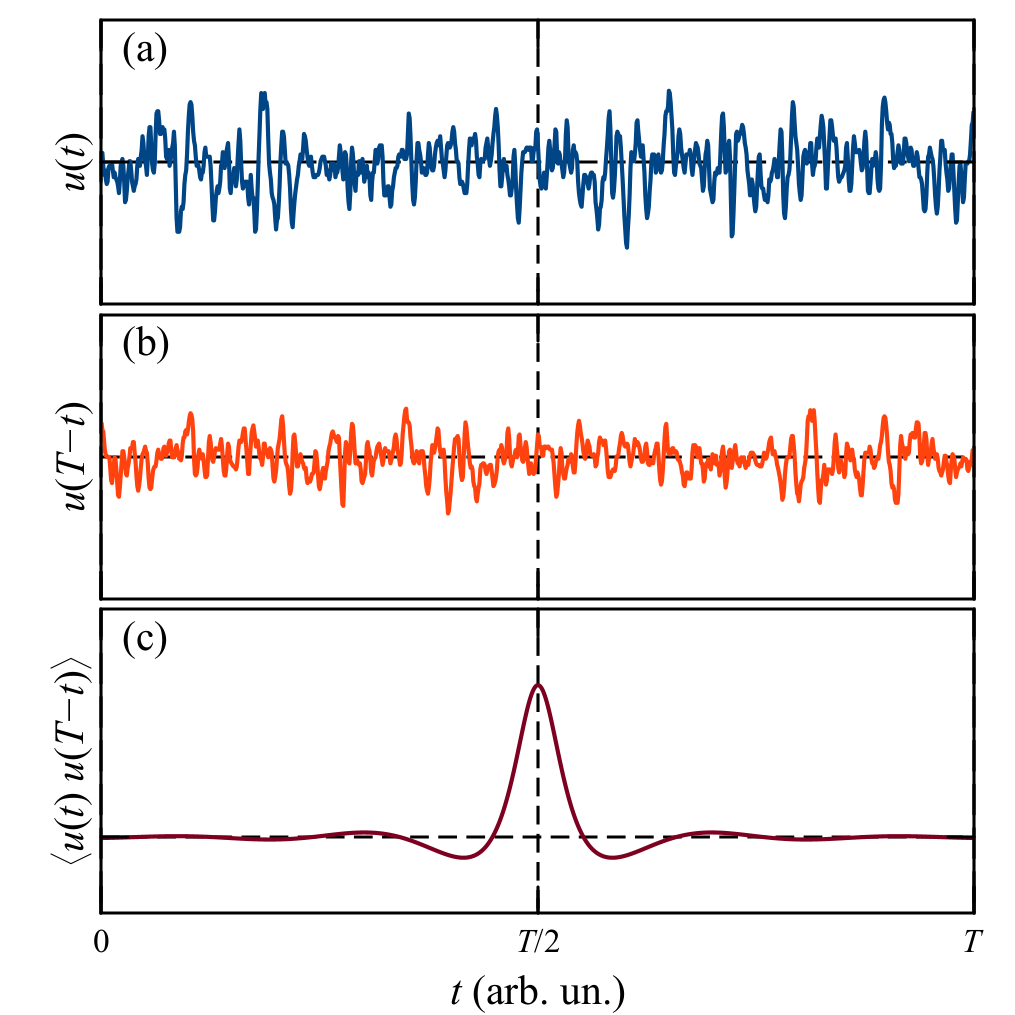}
\caption{Getting correlation function via the time reversal. (a) A fragment of the detected noise signal. (b) The same fragment reversed in time. (c) Expected autocorrelation function of the signal, which can be obtained by multiplication and averaging over multiple realizations of noise traces.
}\label{fig2}
\end{figure}

Schematically, realization of this idea looks as follows.
Consider again some particular fragment of the noise signal $u(t)$, $0<t<T$, detected in a spin-noise experiment [Fig.\ \ref{fig2}(a)]. 
Let us renumber entries of this noise trace in a reverse way.
The data obtained in this way depicts the same signal with exactly reversed sequence of events [Fig.\ \ref{fig2}(b)]. 
By its appearance, this trace does not differ from that shown in Fig.\ \ref{fig2}(a), and it is impossible to say, which of them corresponds to the real time-evolution of the signal, provided that the state of the system is stationary \cite{noisereversal}. 
One can notice, however, that the values of the signal at the center of the fragment ($t = T/2$), for these two traces, are always the same, or, in other words, are totally correlated. 
Thus, we see that the cross correlation function $\langle u(t)u(T-t)\rangle$ of these two signals should be necessarily peaked at the center of the chosen time interval, with its shape being governed by correlation properties of the noise under study [Fig.\ \ref{fig2}(c)].

Therefore, qualitatively, it seems evident that this experimental approach allows one to obtain correlation function of the spin-system noise or, more generally, correlation function of the detected noise signal. 
In contrast to standard procedures of correlation measurements, we perform averaging over realizations with a wide range of time delays, rather than over time delays for a single realization.

Since the function $u(t)$ is a realization of a {\it stationary} stochastic process, its correlation function $K(t,t')\equiv \langle u(t)u(t')\rangle$ depends only on difference of its arguments $K(t,t')\rightarrow K(t-t')$. 
This is why, the function $\langle u(t)u(T-t)\rangle$ measured in our experiment directly corresponds to the standard correlation function $K(\tau)$, where $\tau\equiv T-2t$ and $ -T<\tau<T$. 
Using this correlation function, one can obtain the noise-power spectrum $\mathcal{P}(\nu) = \int e^{-\mathrm{i}2\pi\nu\tau}K(\tau) d\tau$ 
detected in the conventional spin-noise spectrometers. 

All the above treatment was performed in application to SNS whereas the described procedure can be applied to any optical field regardless of origination of its intensity fluctuations. 
These measurements do not require any polarization sensitivity of the photodetector, and the needed noise trace can be directly obtained at the output of the photodetector illuminated by the light under study. 
Using the procedure described above, we will generally obtain the correlation function of the light-intensity noise. 
In the absence of any excess noise, the measured correlation function will contain only a single $\delta$-wise peak corresponding to the ``white'' shot-noise of the photodetector. 
In reality, this peak will have a finite width determined either by the digitization step  $\delta t$, or by the inverse bandwidth of the photodetector.  
Note that these two cases can be distinguished by applying the above procedure to the two output beams (rather than to the same beam). 
In this case, the $\delta$-wise peak related to the Poissonian shot noise will vanish, while all kinds of the excess intensity noise will remain the same\ \cite{probability, probability2}.   
This simple procedure allows one to get rid of the shot-noise contribution to the spin-noise correlation function.

\begin{figure}[t]
\includegraphics[width=\columnwidth]{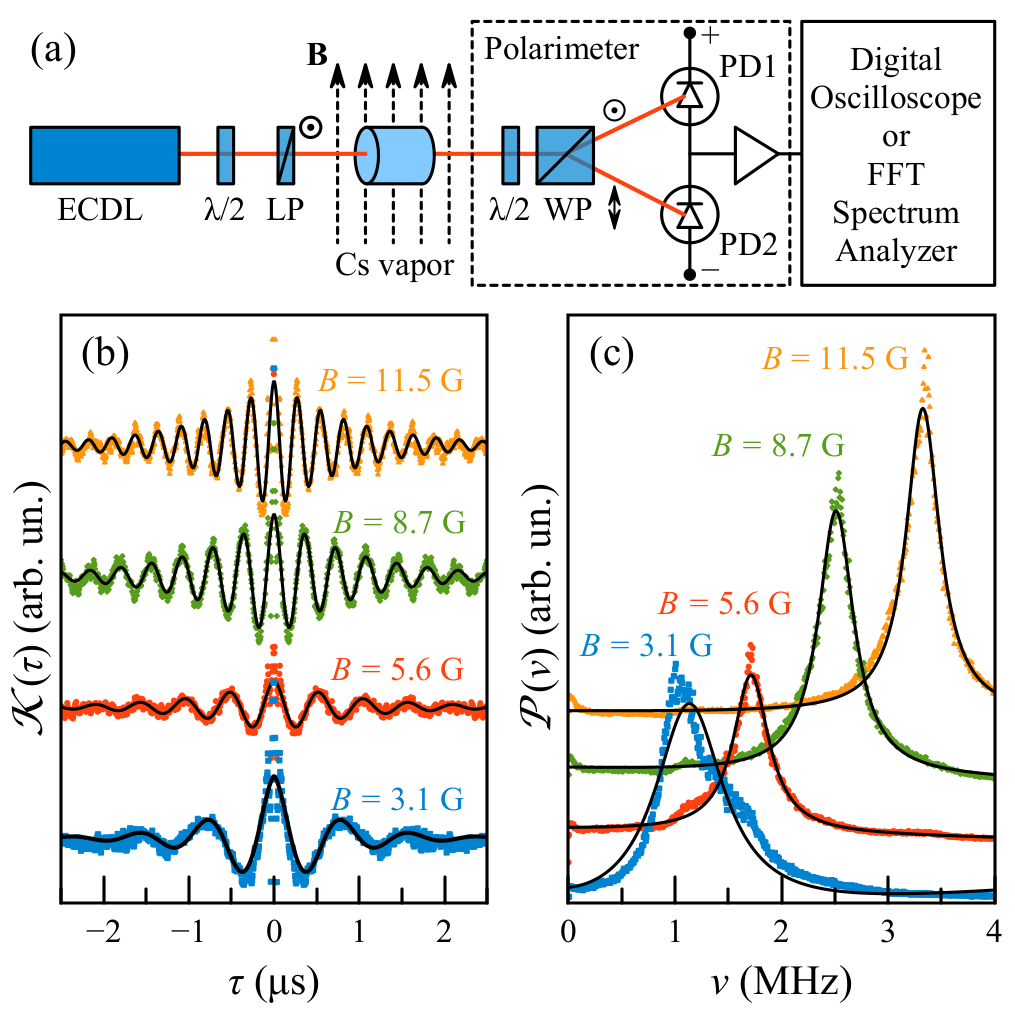}
\caption{
(a)~Schematic of the experimental setup. (b)~Spin-noise correlation functions, defined by Eq.\ \eqref{eq_K}, measured using digital oscilloscope (points) and their fitting by Eq.\ \eqref{eq_Kfit} (black curves). (c)~The spin-noise power spectra measured for different magnetic fields using commercial FFT spectrum analyzer (points) and their fitting by Eq.\ \eqref{eq_Pfit} (black curves).
}\label{fig_exp}
\end{figure}

Experimental arrangement used in this work did not differ essentially from that employed in conventional spin-noise measurements, as shown in Fig.\ \ref{fig_exp}(a). 
Specific features of the setup were related to the procedure of the data processing.
We used an external-cavity diode laser (ECDL) tuned  $\simeq0.5$\ GHz below the $F=4 \leftrightarrow F'=3,4,5$ transition of the Cs D2 line at the wavelength $852.35$~nm\ \cite{hompaper, anom}.

The linearly polarized laser beam, $2$\ mm in diameter and $\sim 2$\ mW in power, with its intensity controlled by combination of a half-wave plate ($\lambda/2$) and a linear polarizer (LP), shined the cylindrical gas cell ($20\times\oslash20$ mm) held at a temperature of 62~\textcelsius.
The magnetic field, $\mathbf B$, applied across the beam (Voigt geometry) was created by a pair of Helmholtz coils.
The noise signal was detected using a standard polarimetric scheme comprised of an additional $\lambda/2$, a Wollaston prism (WP), and a balanced photodetector with two p-i-n photodiodes (PD1 and PD2) ($100$\ MHz bandwidth). The output of the photodetector could be connected either to a digital oscilloscope, which was employed as an analog-to-digital converter, or to a commercial fast Fourier transform (FFT) spectrum analyzer used to compare the results of the correlation and standard spectral measurements.

The continuous noise trace acquired by the oscilloscope, was sliced into multiple $T = 5$\ $\mu$s long segments.
Each data segment digitized with the sampling period $\delta t = 5$\ ns, represented a numbered set (column) of data $u_n^{(k)} \equiv u( kT + n\delta t)$ where $n = 0, 1, \ldots, N$, $N = T/\delta t$, with $N$ being total number of entries in the $k$-th data segment. 
The column of data $\{u_n^{(k)}\}$, multiplied by its reversed copy $\{u_{N-n}^{(k)}\}$, was then averaged over multiple realizations and normalized, thus giving the correlation function 
\begin{equation}
\mathcal{K}(\tau_n) \equiv \frac{\langle u_n^{(k)} u_{N-n}^{(k)} \rangle_k}{\langle|u_n^{(k)}|^2\rangle_k}
\label{eq_K}
\end{equation}
where $\tau_n = (N -2n)\delta t$, and $\langle\ldots\rangle_k$ denotes cumulative moving average over multiple data segments. 

The results of such a treatment are shown in Fig.\ \ref{fig_exp}(b).
As seen from the figure, the dependences $\mathcal{K}(\tau)$ thus obtained represent even oscillating functions with decaying amplitudes. 
The presence of any broadband (``white'') noise, including the shot noise, is revealed as the $\delta$-wise peak at $\tau=0$ and may be ignored.

For comparison, we present, in Fig.\ \ref{fig_exp}(c), the spin-noise spectra of cesium recorded using a standard FFT spectrum analyzer. 
In each magnetic field, these spectral measurements were performed simultaneously with those presented in Fig.\ \ref{fig_exp}(b), so that the experimental conditions, in these two cases, were identical.

The spin-noise correlation functions [Fig.\ \ref{fig_exp}(b)] and spin-noise spectra [Fig.\ \ref{fig_exp}(c)] were fitted using the exponentially damped oscillation and the Lorentzian functions, respecively, with the same fitting parameters: the spin relaxation time, $\tau_s$, and the Larmor frequency, $\nu_L$.
The fitting function for Eq.\ \eqref{eq_K} reads
\begin{equation}
\mathcal{K}(\tau) \approx \mathcal{K}_0 \exp(-|\tau|/\tau_s) \cos (2\pi\nu_L\tau),
\label{eq_Kfit}
\end{equation}
and the fitting function for the spin-noise power-spectrum is a Fourier transform of the function $\mathcal{K}(\tau)$, Eq.\ \eqref{eq_Kfit}\ \cite{sn2}
\begin{equation}
\mathcal{P} (\nu) \approx \sum_{\pm\nu_L} \frac{\mathcal{P}_0/(2\pi\Gamma)}{1 + (\nu\pm\nu_L)^2 / \Gamma^2},
\label{eq_Pfit}
\end{equation}
with $\Gamma$ being the half-width at half-maximum  of the Lorentzian function and $\tau_s = (2\pi\Gamma)^{-1}$.
As expected, in each magnetic field, the spin-noise correlation function and the spin-noise spectrum are perfectly described by the same parameters $\tau_s$ and $\nu_L$\ \cite{Note1}. 

Thus, we demonstrate experimentally that both approaches provide the same information about dynamics of the spin-system. 
However, the most essential computational bottleneck in the spectral decomposition of the data is veiled in the multiple calculations of the FFTs for each data segment.  
To expand the bandwidth of the real-time signal accumulation, the specially configured processor, e.g., based on the field-programmable gate array\ \cite{crookerPRL10}, is required. 
The computation complexity of the presented time-reversal algorithm is much less resource-consuming than FFT analysis, which requires $\mathcal{O}(N \log_2 N)$ operations to obtain each Fourier spectrum\ \cite{fftcalculus}.
The calculation of the correlation function in time domain requires only the matrix multiplication\ \cite{matrixmultiplication}, demanding $\mathcal{O}(N^{1+\varepsilon})$ operations where $\varepsilon\gtrsim0$.
The rest of the algorithm, including averaging procedure, remains the same for the FFT-based and the time-reversal processing.
Thus, we believe that the proposed method, characterized by lower computational expenses and simpler technical realization will make the spin-noise investigations more practical and easier available for researchers in this field of experimental physics. 


To conclude, we have shown that the autocorrelation function of a stationary noise signal can be easily obtained by combining noise-traces with their time-reversed copies. 
The proposed approach retains all the merits of the Fourier spectroscopy but implies the use of cheapper equipment and offers a cost-efficient  computation algorithm. 
The method is  successfully applied to spin-noise measurements in cesium vapor. 
We also show that the proposed idea of the time reversal, being implemented in a two-beam geometry, provides abilities to study the nature of the light-intensity noise and to get rid of the shot-noise contribution to the noise correlation function. 
We believe that the proposed expedient, which significantly simplifies the data processing in the noise spectroscopy, will have a strong impact on further development of the spin-noise spectroscopy. 

\vspace{1.0em}


The authors thank D.\ S.\ Smirnov for fruitful discussions and 
acknowledge Saint-Petersburg State University for the research project 122040800257-5.
The work was carried out at the site of the Resource Center ” Nanophotonics“ of the Research Park, St. Petersburg University.

\end{document}